\documentclass[preprint]{aastex}

\newcommand{\beq}{\begin{equation}}
\newcommand{\eeq}{\end{equation}}

\newcommand{\hii}{H~{\sc ii}~}
\newcommand{\kms}{\mbox{ km s$^{-1}$~}}

\newcommand{\cmc}{\mbox{ cm$^{-3}$~}}

\shorttitle{Ultracompact to Extended HII Regions}
\shortauthors{Garcia-Segura, Franco \& Kurtz}

\begin{document}

\title{From Ultracompact to Extended \hii Regions. II : Cloud Gravity and
Stellar Motion}

\author{Jos\'e Franco\altaffilmark{1}}
\affil{Instituto de Astronom\'{\i}a-UNAM, Apdo Postal 70-264, 04510
M\'exico D.F., Mexico}
\author{Guillermo Garc\'{\i}a-Segura\altaffilmark{2}}
\affil{Instituto de Astronom\'{\i}a-UNAM, Apdo Postal 877, 22800 Ensenada,
Baja California, Mexico}
\author{Stanley E. Kurtz\altaffilmark{3}}
\affil{Centro de Radioastronom\'{\i}a y Astrof\'{\i}sica-UNAM,
Apdo Postal 3-72 (Xangari), 58089 Morelia, Michoac\'an, Mexico}
\altaffiltext{1}{Email address: pepe@astroscu.unam.mx}
\altaffiltext{2}{Email address: ggs@astrosen.unam.mx}
\altaffiltext{3}{Email address: s.kurtz@astrosmo.unam.mx}

\begin{abstract}
  
  The dynamical evolution of \hii regions with and without stellar motion in
  dense, structured molecular clouds is studied. Clouds are modeled in
  hydrostatic equilibrium, with gaussian central cores and external halos that
  obey power laws as $\rho \propto r^{-2}$ and $\rho \propto r^{-3}$. The total
  pressures at the core centers are $P/k = 10^9 $~K\cmc. The cloud gravity is
  included in a simplified manner as a time-independent, external force.
  Stellar velocities of 0, 2, 8, and 12 \kms are considered; all are less than
  the equilibrium dispersion velocity of the cloud (13~\kms). When stellar
  motion is included, stars move from the central core to the edge of the 
  cloud, producing transitions from ultracompact to extended \hii regions as
  the stars move into lower density regions. The opposite behavior is also
  feasible when stars move toward the cloud cores. The main conclusion of our
  study is that ultracompact \hii regions are pressure-confined entities while
  they remain embedded within dense cores. The confinement comes from either
  ram or ambient pressures, or a combination of both. The survival of
  ultracompact regions (by definition $\lesssim 0.1 $ pc) depends on the position
  of the star with respect to the core center, the stellar life-time, and the
  crossing time of the cloud core. Stars with velocities less than the cloud
  dispersion velocity can produce cometary shapes with sizes under
  0.1 pc at evolutionary times of $2 \times 10^4 $ yr or greater, in 
  statistical agreement with observations. The sequence Ultracompact \hii 
  $\rightarrow$ Compact \hii $\rightarrow$ Extended \hii shows a rich variety
  of unpredictable structures due to ionization-shock front instabilities.
  Some ultracompact \hii regions with a core-halo morphology could be 
  explained by self-blocking effects, when stars overtake and ionize leading, 
  piled-up clumps of neutral gas.  Because of computational restrictions, we
  use thermal energy to support the cloud against gravity; the results remain 
  the same if other types of isotropic cloud support are used,
  such as hydrodynamic or turbulent pressure. 

\end{abstract}


\section{Introduction} 

The classical picture of \hii regions presents the idea of a newly-formed
massive star that begins to produce large quantities of UV photons and hence 
to ionize the surrounding medium. If the star is still embedded within the 
molecular cloud from which it formed, then the molecular gas will dissociate 
and ionize, and an ionization front will form, expanding rapidly outward to 
form the initial Str\"omgren sphere. The ionization front expands very rapidly
at first, but soon slows, and approaches the sound speed after a few
recombination times. The newly ionized gas will be greatly over-pressured with 
respect to the original molecular gas and, in the classical picture, will 
expand toward pressure equilibrium with its environment. In pressure
equilibrium, for ionized gas velocity dispersion $c_i$ and initial (molecular) 
cloud velocity dispersion $c_0$, the final radius can be expressed in terms of the 
initial Str\"omgren radius, $R_s$, as

\begin{equation}
 R_f = R_s \biggl( {c^2_i\over c^2_0}\biggr)^{2/3}\ .
\end{equation}

\noindent
For typical values of $c_i$ and $c_0$, one expects the final radius to be 
nearly 100 times its initial value. The strong shock approximation for the 
expansion gives the Str\"omgren sphere radius as a function of time (Spitzer 
1954)
\begin{equation}
R_{HII}\simeq R_S \left(1+ \frac{7}{4} \frac{c_i t}{R_S}\right)^{4/7} \ .
\end{equation}

\noindent
Adopting $c_i = 10$~km~s$^{-1}$ as the sound speed, and assuming a
factor of 100 expansion, implies that the \hii region will reach
pressure equilibrium in a time longer than 3 $\times 10^6$ years.
(Equation 2 provides an upper bound because the dynamical evolution
departs from the strong shock approximation as the expansion velocity
decreases.) At this time it will have a radius of about one parsec.
The basic physical processes relevant for the detailed structure and
expansion of \hii regions are discussed extensively in Osterbrock
(1989). Other important considerations driving the formation and
expansion phases are discussed in Franco, Tenorio-Tagle \& Bodenheimer
(1990). The salient point here is that the general scheme outlined
above for the expansion of recently formed \hii regions is
problematic, in that it predicts that the region will rapidly expand
out of its initial, small, dense (ultracompact) state. In particular,
adopting 0.1~pc as the nominal size for an ultracompact (UC) \hii
region, one expects the lifetime of this phase to be of order
$r/c_{i}$, or about $10^4$~yr. Many more UC~HII regions are seen than
this naive view predicts, hence the conclusion is that their lifetimes
must be longer than anticipated by this simple theory.

During the past decade various attempts have been made to address this 
so-called ``lifetime problem'' (see Garay \& Lizano 1999, Kurtz et al. 2000, 
and Churchwell 2002 for reviews). These attempts can be divided into two 
categories: those which prolong the life by providing a confining force, and 
those which extend the life by adding a mass source, thus replenishing the 
ionized gas. In both cases, however, the net result of the process is that the
pressure of the ionized region is increased, leading to densities and sizes 
comparable to those observed in UC~\hii regions for extended periods of time.

Attempts to prolong UC~\hii region lifetimes by providing an
additional confining force include the bow-shock model (Van Buren \&
McCray 1988; Van Buren et al. 1990; Mac Low et al. 1991; Van Buren \&
Mac Low 1992), using thin-shell solutions for the bow-shock
computation. These models concluded that UC~\hii regions could be
confined by ram pressure when they travel through a cloud.
Garc\'{\i}a-Segura \& Franco (1996, hereafter, paper~I) showed that
other sources of pressure may also play an important role. For
example, thermal pressure within a cloud core can itself confine a
UC~\hii region to small sizes, as suggested by De~Pree, Rodr\'\i guez
\& Goss (1995).  It seems unlikely, however, that the simple action of
thermal pressure can give rise to the variety of observed
morphologies, and additional ways of increasing the total pressure
have been explored. For instance, one can achieve this by replenishing
the ionized gas.  Hollenbach et al.(1994), Yorke \& Welz (1996), and
Richling \& Yorke (1997) modeled UC~\hii regions resulting from the
photo-evaporation of circumstellar disks, while Dyson, Williams, \&
Redman (1995) and Lizano et al. (1996) modeled them as
photo-evaporating clumps. A key result of all these studies is the
realization that the ``compactness'' (i.e., the small diameter and
high density) of the HII region may not be a good indicator of its
youthfulness.

Following along these lines, in this paper we discuss models that
include stellar motions, and combine the effects of ram and thermal
(or other) pressures. Again, as in Paper I, we use thermal pressure as
an illustration of the impact that {\it any} isotropic pressure may have 
on the dynamical evolution of \hii regions.  Here we present gas
dynamical simulations that complement those of paper I. The inclusion
of stellar motion introduces a preferential direction for the pressure
variation and gives rise to a variety of different situations as the
stars leave their parental cloud cores and evolve into extended \hii
regions. A wide variety of shapes is encountered for extended \hii
regions, and we confirm the appearance of saw-tooth ionization-shock
fronts discussed in paper~I and Williams (2002).

Another aspect considered in this paper is that \hii regions are
calculated off-center from their parental clouds. Previous studies
(i.e., Franco et al. 1990 and references therein) considered stars
located at the peak of the density structures, and the solutions for
the expansion assumed spherical symmetry from the core center. In the
off-center cases, and for the core densities used here ($\sim 10^7$
\cmc), a large number of ionizing photons is required for complete
core ionization and our simulations admit the possibility that the
outer core material either remains un-ionized or recombines at a late
evolutionary stage.

Section 2 gives an observational background from recent radio studies. \S 3 
describes the model and the assumptions made. \S 4 explains and shows the gas 
dynamical simulations, and \S 5 discusses the results and gives our conclusions.

\section{Observational Background}

The physical properties of \hii regions span orders of magnitude in scale. The
classes most closely linked to star formation are the smallest, densest, and 
(possibly) youngest stages: the compact, ultracompact, and hypercompact HII 
regions. Compact \hii regions were first identified as such by Mezger et al. 
(1967), who characterized them as having sizes from 0.06 to 0.4~pc and electron 
densities around 10$^4$~cm$^{-3}$. The smallest and densest of these came to be
known as ultracompact \hii regions. By the mid-1980s around a dozen of these
UC~\hii regions were known. The situation changed dramatically with the survey 
of Wood \& Churchwell (1989) which identified over 70 UC~\hii regions. Further 
surveys carried out by Becker et al. (1994), Kurtz, Churchwell \& Wood (1994), 
Walsh et al. (1998), and Giveon et al. (2005) identified many more, and at 
present the number of probable ultracompacts is of order 1000. These regions 
are (observationally) defined as having sizes $\lesssim$ 0.1~pc and densities 
$\gtrsim 10^4$~cm$^{-3}$. The surveys revealed that UC~\hii regions present a
variety of morphological types, including cometary, core-halo, shell,
spherical, and multiply-peaked (see Wood \& Churchwell 1989, Fig. 143). A 
modification to this morphological scheme, including the addition of a bipolar
morphology, has been suggested by De Pree et al. (2005). Hypercompact \hii 
regions, only recently recognized as a distinct class (e.g., Sewilo et al. 
2004), are even smaller and denser than UC~\hii regions, but so far are fairly
rare, with only about a dozen presently known.

Most of the radio surveys mentioned above suffered from the observational bias 
that they were insensitive to regions larger than about $30''$. This bias 
arises from the nature of interferometric observations, which are sensitive 
only to a range of angular sizes, determined by the range of baseline lengths. 
Although some lower resolution studies exist (e.g., Garay et al. 1993), the larger
UC~\hii region surveys were made with arc second or better angular
resolution. For the radio interferometers used in these surveys (the VLA and the
ATCA) this implies observations that are insensitive to structures larger than 
approximately 30$''$.

Low resolution observations of UC~\hii regions indicate that this observational
bias may result in a significant misrepresentation of some regions. Kurtz et 
al. (1999) and Kim \& Koo (2001) made low resolution VLA observations of UC~\hii
regions from the Kurtz et al. (1994) and the Wood \& Churchwell (1989) surveys,
respectively, and found that a large fraction had extended free-free emission
surrounding the ultracompact components. The physical relationship between the
extended and the compact components is unclear; the evidence for a relationship 
is morphological in most cases. Kim \& Koo report radio recombination line
observations which support the hypothesis of a relation based on close
agreement of the line velocities between the ultracompact and extended
components.

The presence of the extended component surrounding the ultracompact
component has been known since the seminal work of Mezger et al.
(1967). Since that time attention has focused on the ultracompact
component and relatively little has been done to explore what role the
extended gas plays in the dynamics of the UC~\hii regions. Further
observational work is needed in this area to confirm the relationship
between the compact, high density gas and the extended, low
density gas. If confirmed, there will be a number of
ramifications for our understanding of UC~\hii regions. First, the
stellar content of the \hii regions (i.e., the ionizing photon flux)
will have to be revised upward to account for the much greater ionized
mass. Second, the theoretical models currently proposed will have to
account for the extended emission. Although the best-known HII region,
Orion, has ionized gas ranging from $n_e$ of about 400~cm$^{-3}$ to
over $10^5$~cm$^{-3}$ (Felli et al. 1993); none of the currently
proposed UC~HII models can easily explain the coexistence of such
different densities.

A general scheme that could give rise to a wide range of ionized gas
densities was suggested by Franco et al. (2000), and shown pictorially
in Figure 8 of Kim \& Koo (2001). Using the free-free radio-continuum
spectral index of three UC and hypercompact \hii regions, Franco et
al. (2000) found that the ionized gas emission can be fit by power-law
density distributions with exponents between $-2$ and $-4$. These are
rather steep gradients, the more extreme cases being significantly
steeper than the gradients found in molecular line studies of
star-forming clouds (which show exponents between $-1.5$ and $-2$,
typical of equilibrium isothermal clouds). The reasons for this
behavior are unclear.  As Franco et al. point out, the extreme value
derived (close to $-4$) may be an over-estimate of the actual density
gradient.  Despite our lack of understanding of the nature and
maintenance of such steep gradients, here we consider clouds with
power-law exponents of $-2$ and $-3$.  The suggestion, then, is that
the density gradients and range of structured scale sizes in molecular
clouds may give rise to compact and extended structures when ionized
by the nascent stars.  The present article is an attempt to develop
this model in a quantitative way.

\section{Initial considerations} 

We assume a cloud consisting of an internal core of radius $r_{\rm c}$ and a
halo in hydrostatic equilibrium, with a constant total velocity dispersion ($ 
P\, = \, \rho \, c_{\rm 0}^2 $, and $c_{\rm 0} \, = \, {\rm constant}$),
\beq
\vec{\nabla}  P\,\,  = \,\, - \,\, \rho \,\, \vec{g} \,\, \Rightarrow
\,\, \vec{\nabla} \rho\,\,  = \,\, - \,\, \frac{\rho}{c_{\rm 0}^2}\,\,  
\vec{g} \,\,.
\label{eqn:gradP}
\eeq

\noindent
For simplicity, due to computational restrictions, we assume that the gas is 
always described by a single equation of state and the total velocity
dispersion is set equal to the isothermal sound speed, $c_{\rm s}$. Thus, the
resulting {\it effective} temperature is representative of the kinetic energy
required to provide support against gravity, and is well above any actual 
molecular cloud temperature (i.e., the contribution from turbulent pressure
is included in an implicit manner). Assuming that the cloud is spherically 
symmetric ($\vec{\nabla} \longrightarrow \frac{\rm d}{{\rm d r}}$ and $\vec{g} 
\longrightarrow  g_{\rm r}$), we solve only along the radial coordinate:
\beq
\frac{\rm d \rho}{{\rm d r}} \,\, = \,\, - \,\, \frac{\rho}{c_{\rm s}^2}\,\,
g_{\rm r} \,\,.
\label{eqn:drho}
\eeq 

For the first set of simulations (designated model A) we assume that the halo 
density falls off as an $r^{-2}$ power law, as in a self-gravitating isothermal
cloud:
\beq
\rho (r) \,\, = \,\, \rho_{\rm c} \,\, \left(  r / r_{\rm c} \right)^{-2}
 \,\,\,\, {\rm for} \,\,\,\,  r \geq r_{\rm c} \,\,.
\label{eqn:rho-halo}
\eeq 

\noindent
Solving for $g_{\rm r}$ with equations {\ref{eqn:drho}} and {\ref{eqn:rho-halo}}, we find
\beq
g_{\rm r} \,\,= \,\, \frac{2 \,\, c_{\rm s}^2}{r} \,\,= \,\,
\frac{2 \,\, c_{\rm s}^2}{r_{\rm c}} \,\,\left(  r / r_{\rm c} \right)^{-1}
 \,\,\,\, {\rm for} \,\,\,\,  r \geq r_{\rm c} \,\, .
\eeq

Inside the core, $g_{\rm r}$ must grow from zero up to $ 2 \, c_{\rm s}^2 / 
r_{\rm c}$. A simple way to achieve this is to let $g_{\rm r}$ grow linearly 
within the core, as $ g_{\rm r} \, = \, A \left( r/r_{\rm c} \right)$. In order
to join the halo solution (equation {\ref{eqn:rho-halo}}), we require that 
$ A \, = \, 2 \, c_{\rm s}^2 / r_{\rm c} $.

The density distribution inside the core is found by integrating equation 
{\ref{eqn:drho}} with the new $g_{\rm r}$, giving
\beq
\rho (r) \,\, = \,\, \rho_0 \,\, {\rm exp} \left[ - \left(  r / 
r_{\rm c} \right)^2
\right]  \,\,\,\, {\rm for} \,\,\,\,  r \leq r_{\rm c} \,\,,
\label{eqn:rho-core}
\eeq
where $ \rho_0 $ is the central density at $r=0$, and $ \rho_{\rm c} = 
\rho_0 / {\rm e} \,$ .

In summary, the density distribution of the cloud is given by 
\beq
\rho (r) \,\, = \,\, \cases { \rho_0 \,\, {\rm exp} \left[ - \left(  
r / r_{\rm c} \right)^2 \right] & {\rm for}  $r \leq r_{\rm c}$
\cr 
\rho_0 / {\rm e} \,\, \left(  r / r_{\rm c} \right)^{-2} & {\rm for}
 $ r \geq r_{\rm c} $ \cr }
\label{eqn:A-rho}
\eeq
and the gravitational acceleration by
\beq
g_{\rm r} \,\, = \,\, \cases { 2 \,\, c_{\rm s}^2 / r_{\rm c} \,\,\,\, 
\left(  r / r_{\rm c} \right)  & {\rm for}  $r \leq r_{\rm c}$
\cr
2 \,\, c_{\rm s}^2 / r_{\rm c} \,\,\,\, 
\left(  r / r_{\rm c } \right)^{-1}  & {\rm for} $ r \geq r_{\rm c} $ .\cr }
\eeq

For the second set of simulations (designated model B) we assume that the halo
density falls off as an $r^{-3}$ power law, and using the above approach we 
find the density distribution
\beq
\rho (r) \,\, = \,\, \cases { \rho_0 \,\, {\rm exp} \left[\,\, - 3/2 \,\, \left(
r / r_{\rm c} \right)^2 \right] & {\rm for}  $r \leq r_{\rm c}$
\cr
\rho_0/{\rm e}^{3/2} \,\, \left(  r / r_{\rm c} \right)^{-3} & {\rm for}
 $ r \geq r_{\rm c} $ \cr }
\label{eqn:B-rho}
\eeq
and the gravitational acceleration
\beq
g_{\rm r} \,\, = \,\, \cases { 3 \,\, c_{\rm s}^2 / r_{\rm c} \,\,\,\,
\left(  r / r_{\rm c} \right)  & {\rm for}  $r \leq r_{\rm c}$
\cr
3 \,\, c_{\rm s}^2 / r_{\rm c} \,\,\,\,
\left(  r / r_{\rm c } \right)^{-1}  & {\rm for} $ r \geq r_{\rm c} $. \cr }
\eeq

For both sets of simulations we assume that the star was born $in$ $situ$,
inside the core of its parental cloud (modeling of ``run-away'' stars is not 
considered). The largest expected stellar velocity that such a cloud can 
produce is given by the dispersion velocity in hydrostatic equilibrium. Noting 
that (see \S 2.1 in paper I) $P\,= \, 16/9 \,\, \pi \, G \,\rho_{\rm c}^2 \,
r_{\rm c}^2 \,= \, \rho_{\rm c} \, c_s^2 $, and solving for $c_s$, we find
\beq
c_s \,\,= \,\, 4.07 \,\, r_{0.1} \,\,n_6^{1/2} \,\, \kms
\eeq
where $r_{0.1}$ is the core radius in units of 0.1 pc and $n_6$ the core 
density in units of $10^6$~\cmc. Thus, for core densities of $10^7$ \cmc and 
radii of 0.1~pc, stellar velocities up to $\sim 13$ \kms can be considered.

\section{2-D gas dynamical simulations}  

The numerical simulations are performed with the gas dynamical MHD code ZEUS-3D
version 3.4 (Stone \& Norman 1992; Clarke 1996; see paper I for details). We 
use Cartesian coordinates with the Y-axis being the symmetry axis (i.e., a slab 
geometry). Thus, the star can be placed anywhere in the X-Z plane. This is a 
particularly safe choice because it does not introduce any axis artifact in the 
computations (as would cylindrical coordinates). 

The set-up is similar for all models: the star is fixed on the computational 
mesh and remains at the same location during the computation. The stellar 
motion is simulated by setting the gas to a single speed throughout the mesh. 
As the simulation proceeds, the outer boundary of the Z-axis is updated with 
the incoming gas at the stellar velocity. This update depends on the nature of 
the problem and accounts for the change in the cloud density 
distribution as a function of position. In reality, the situation is far more 
complicated because the velocity is also a function of time: the stellar velocity changes
as it moves through the gravitational potential of the cloud but, unfortunately,
this variation cannot be included here in a self-consistent manner.  All models 
have the same numerical resolution of 250 zones along both the X and Z axes.

We use the same approach as paper I to model the \hii region (see Bodenheimer, 
Tenorio-Tagle \& Yorke 1979), i.e., we solve a radial integral to find the 
position of the ionization front (Str\"omgren 1939). The temperature inside the
\hii region is set to $10^4$ K, the approximate photoionization 
equilibrium temperature.

The modeling of a ``realistic" UC~\hii region includes a hypersonic
stellar wind (i.e., an ultracompact bubble) and is very expensive in
computational time, owing to the fact that the wind speeds from
massive main sequence stars are of order $10^3$ \kms when their bubble
sizes are of order $10^{-2}$ pc. This forces the Courant condition to
calculate very small time-steps during the simulations.  Thus, it is
possible to include stellar winds in 2-D simulations that cover
physical times of order $10^3$ yr, but this is very time consuming
when the required physical times are of order $10^5$ yr. For this
reason, we do not include the effects of the stellar wind in the
present models. Nonetheless, Paper I showed that at these high
densities the size of the UC~\hii region is primarily determined by
the gas pressure. Thus, our results can be considered qualitatively
correct.

To begin our study we computed several cases (not shown here) in which the star 
was located at the center of the core (given by equation 8) and had zero velocity. 
For a core radius of 0.1 pc, a central density of $10^7$ \cmc and an ionizing 
photon flux of $F_{\star} = 10^{48}$ s$^{-1}$, the models followed similar 
tracks as those described in the 1-D solutions of Paper I. This is 
understandable, because the gaussian core described by equation 8 produces a 
central plateau which resembles the constant density medium used in the first 
part of Paper I. Thus, pressure equilibrium is achieved on timescales matching 
those of Paper I. The novel aspect here is the inclusion of gravity, which 
affects the final density structure. This is illustrated in the next paragraph.

A second set of models, also without stellar motion (models A0 and B0; see 
Table 1), are calculated with the star at the edge of the core (Figure 1a). The
purpose of these models is to study the expansion and dynamics of the \hii 
regions in the two different density ramps given by equations {\ref{eqn:A-rho}
and {\ref{eqn:B-rho} for $r \geq r_{\rm c}$. Model A0 is shown in detail in 
Figure 1b, while model B0 is shown in Figure 1c. The ionized gas has a flat 
density distribution during the first forty thousand years (first two curves) 
in both cases. But as the evolution proceeds, the ionized gas -- subject to the 
cloud gravity -- adjusts its density distribution to come into hydrostatic 
equilibrium. This is clearly seen in model A0, where pressure equilibrium is
achieved (curves 7 -- 10) and the final density distribution follows the initial
density ramp. The final density is a factor of 200 lower, however, because the 
gas was originally at $10^2$~K but hydrostatic equilibrium is achieved at 
$10^4$~K. In contrast, model B0 does not reach pressure equilibrium in the
computational domain, and a blister-like region or champagne flow 
(Tenorio-Tagle 1979) is produced. Gas is permanently photo-evaporated from the 
cloud, resulting in a steady-state mass loss. As a result, the final solution 
is closer to an $r^{-2}$ profile than to the original $r^{-3}$ density ramp.

We note that the density ramps for these models are off-center from
the stellar position; i.e., the origin of the power law does not
coincide with the stellar coordinates. In addition, in some models,
the star moves. As such, the spherically symmetric solutions given by
Franco et al. (1990) are not directly applicable in all cases.

A third set of models (A2, A8, A12 and B2, B8, B12) include stellar motion at 
2, 8, and 12 \kms (Table 1). In these cases the star moves from the core center
toward the edge, eventually leaving the core and entering the density ramp 
described in equations {\ref{eqn:A-rho}} and {\ref{eqn:B-rho}}. While the star 
is still inside the core, these models create cometary shapes arising from the
leading bow-shock, with sizes typical of UC~\hii regions. When the star enters
the density ramp, the \hii region grows in size along the density gradient and
drops in density. If the stellar velocity is high compared to the \hii 
expansion velocity then the cometary shape is maintained. If the stellar 
velocity is low compared to the \hii expansion velocity, the region quickly
evolves into a blister morphology. The results of this third set of models,
including a constant stellar motion, are shown graphically in Figures 2 -- 6.

Figures 2 and 3 show models A2 and B2, respectively. Owing to the lower stellar
velocity, four time frames are adequate to show the evolution. In these two 
simulations the star is initially at the core center and moves toward $+z$. 
Our computational method leaves the star fixed at the position $[(x,z)_{\star} 
= (0.25,0.1)]$ while the gas moves toward $-z$ at 2 \kms. At the time of the 
first frame ($0.8 \times 10^5$~yr) the star has already reached the core edge 
and entered into the density ramp. In subsequent frames, the star has moved 
outward in the density ramp, leaving the cloud progressively further 
behind (toward $-z$). The behavior of models A2 and B2 is qualitatively 
similar. However, owing to the steeper density gradient, model B2 grows more rapidly, 
showing greater expansion.

Figure 4 shows the final shape of the ionized region of models A0, B0, A2, and 
B2 (the final times shown for each model are different, as described in the 
figure caption). 
All four models present a cometary morphology, but in all cases the
gas dynamics correspond to a champagne flow, not a bow shock. The sizes and
shapes of the ionized regions are different (except for the B0 case) than 
those of the final cavities shown in Figures 1a, 2, and 3. This is because the 
density of the ionized region tends to a uniform value, washing away
the initial gradient and, as a consequence, the ionization front recedes (i.e.,
it becomes a recombination front). Thus, the gas in the outer part of the
expanding cavity recombines and cools. The pressure drops quickly in these
regions, forming blobs and filaments due to thermal and dynamical instabilities.
The effects are clearly noticeable in the B2 model (last frames of Figures 3 
and 4).    

Figure 5 shows the total (5a) and ionized (5b) gas densities for
models A8 and A12. A cometary morphology is clearly evident during the
first half of the evolution for both models. Unlike the A2 model,
for which this morphology arose from a champagne flow, the higher
velocities of the A8 and A12 models result in bow shocks.
During the second half of the evolution, as the star moves into much
lower density gas, the cometary arc opens substantially. Also,
instabilities in both the ionization and shock fronts, as those
described in Paper I and Williams (2002), create complex structures
with finger-like condensations that resemble elephant
trunks (see also Williams et al. 2001). Indeed, it is not clear that
at late times these ionized regions would even be recognized as being
cometary. Corresponding time frames between the A8 and A12 models show
that the \hii region has evolved to a larger size for the A12 model
than the A8 model. This is because the higher stellar velocity of the
A12 model has carried the star to a lower density region of the cloud,
where the \hii region can expand more freely.

Figure 6 shows the total (6a) and ionized (6b) gas densities for
models B8 and B12. Compared to the A8 and A12 models, the B8 and B12
models evolve more rapidly, expanding beyond the ultracompact stage in
about half the time (the time sequence shown in Figure 6 is shorter
than that of Figure 5). As stated in the case of Figure 4, the
ionization front is always inside the expansion cavity, and the gas 
recombines at several locations. Perhaps most
intriguing about the B8 and B12 models is that both pass through a
phase (at 52,000~yr and 40,000~yr, respectively) when distinct high
density components are seen within the HII region. These high density
components have sizes and densities typical of UC~\hii regions, but
they are embedded in larger, lower density ionized gas more typical of
compact \hii regions. This is discussed further in \S 5.4.

The sequence Ultracompact \hii $\rightarrow$ Compact \hii
$\rightarrow$ Extended \hii shows, as stated above, a rich variety of
unpredictable structures due to the I-S front instability (see paper
I). Figures 5-7 show examples of how rich the development of the I-S
front instability can be. The small scale structures found in the
computations (Fig. 7) are always transient. The lifetime of these
structures, or elephant trunks, is proportional to their mass and the
size of the \hii region, and inversely proportional to the stellar
speed. This is because their destruction depends on the
photo-evaporation rate, which in turn depends on the position and
orientation of the ionizing source. The model does not purport to
explain the M16 pillars, which are 50 times larger, but does suggest,
as discussed in more detail by Williams et al. (2001), how their
formation might occur.

\section{Discussion and Conclusions} 

The results presented here show a number of important features that
may be useful in understanding some of the observational puzzles of
\hii region evolution.  Our simulations are performed with simplified
models, however, and further studies with additional refinements are
needed. In particular, the gravitational acceleration is treated here
as a time-independent radial function, and the stellar velocity is
simply set to a constant value. Obviously, both quantities are continually
modified as the cloud is photo-evaporated and the star changes location
in the gravitational field.  Thus, a self-consistent treatment of the
problem will provide better details of the evolutionary features.
Also, we do not include pre-existing clumps in the cloud matter, nor
large-scale vorticity in the gas velocity, which will add complexity to
the resulting structures. Despite these restrictions, our results
provide an adequate qualitative guideline to the effects resulting
from the motion of recently formed massive stars in centrally
condensed cloud cores.

\subsection{The Star Formation Rate}

The results for model A0 indicate that solutions exist in which clouds
are not completely disrupted by ionization fronts; hence, star
formation could proceed for longer times in those clouds. Because
stellar motion is also able to confine \hii regions by ram pressure in
the bow shock, the ionized gas is confined within the cloud and the
star formation rate of the cloud is relatively unaffected. The results
are quite different for model B0, with a steep gradient, or any of the
models with larger stellar velocities, (8~km~s$^{-1}$ or above) in which
the rapid expansion of the ionized gas in the lower density regions of
the cloud would quickly lead to cloud destruction, thus ending the
star formation phase (Franco, Shore \& Tenorio-Tagle 1994;
Diaz-Miller, Franco \& Shore 1998).

\subsection{Density Gradients}

The results for model A0 are also interesting because if pressure
equilibrium is achieved, the density gradient in the \hii region will
match the original density gradient of the parental cloud (see Franco
et al. 2000). However, model B0 shows that molecular cloud density
gradients above $\rho \propto r^{-2}$ cannot lead to pressure
equilibrium, indicating that \hii region gradients obtained in these 
cases probably do not reflect the initial cloud conditions.

Another aspect that is important to stress is that champagne flows are
transient, and in some cases they may be short-lived. Figures 1 -- 4
show that, except for model B0, the initial champagne flow disappears
either because the \hii region reaches pressure equilibrium or the
ionized gas recombines. The latter case leads to neutral outflows
generated by a recombination front, as originally discussed for
disk-like clouds by Franco et al. (1989). Three scenarios give rise to
more-or-less steady champagne flow solutions: density gradients
steeper than $\rho \propto r^{-2}$ at the core edge; ionizing photon
rates high enough to ionize the whole core; and core densities much
smaller than the value used here ($n_0 = 10^7$ \cmc).  The latter
possibility appears to be ruled out for hot molecular cores (which
have densities equal to or even larger than $10^7$ \cmc).

\subsection{The Lifetime Problem and UC~\hii Morphologies}

Any proposed solution to the lifetime problem for UC~\hii regions must
explain how dense, compact ionized gas can remain in that state for
timescales of order $10^5$~yr. We find that model A0 (the stationary model 
with $\rho \propto r^{-2}$, and also the stationary 1-D model of Paper I)
and models A2 and B2, (the low stellar velocity models) are able
to produce UC~\hii regions smaller than 0.1 pc at evolutionary
times larger than $10^5$ yr. An additional important effect, not
considered here, is UV absorption from dust grains inside the cloud
cores (Diaz-Miller et al.  1998; Arthur et al. 2005). This substantially
reduces the size of \hii regions in dense
places, and our results provide only upper limits for the sizes of the
photoionized regions. This makes even easier the confinement of
UC~\hii regions during the required time scales.

A star moving at 1 \kms will leave a 0.1~pc core in $10^5$ yr
(starting from the core center). The reason why solutions with 2 \kms
stellar velocities are still within the range of UC~\hii parameters is
because of ram pressure confinement.  UC~\hii regions remain
ultracompact ($\lesssim 0.1$ pc) as long as they remain inside the cloud
core. In this case, the lifetime as an ultracompact region would be
defined by the core crossing time. In the event that the stellar orbit
never reaches the core edge, the ultracompact lifetime would be the
stellar lifetime, as discussed in Paper I.

Wood \& Churchwell (1989) found that the cometary and core-halo
morphologies were the most common (20\% and 17\%, respectively, in
their sample), suggesting that these morphologies have longer
lifetimes than the other morphologies, thus making their detection
more probable.  For comparison, we find the evolutionary time for each
model when the solution (still dominated by ram pressure) reaches a
size of 0.1 pc. These times (see Table 1) are: A2-B2 $\sim 8 \times
10^4$~yr, A8-A12 $\sim 6.4 \times 10^4$~yr, and B8-B12 $\sim 4 \times
10^4$~yr.  The times for the higher velocity models are roughly
one-third of the nominal $10^5$~yr UC~\hii lifetime, consistent with the 
37\% of the Wood \& Churchwell sample showing a cometary or core-halo 
morphology.

\subsection{The Coexistence of Compact and Extended Emission}

An intriguing self-blocking effect, in which the star overtakes a
leading, piled-up clump of gas and ionizes it, is seen in models B8
and B12. The effect occurs at 52,000~yr and 40,000~yr, respectively,
shown in frames 4 and 11 of figure 6b. Such regions could appear in
radio maps as having a core-halo morphology. Moreover, if the linear
extent of the ionized gas is several tenths of a parsec or more when
the self-blocking occurs, this phenomenon could appear as an
ultracompact \hii region coexisting with extended emission. The \hii
density contrast in frames 4 and 11 is about 20. For comparison, inferred
values for the density contrast in the Orion \hii region range from
about 40 (Wilson \& J\"ager 1987) to nearly 600 (Felli et al. 1993),
with the larger value resulting from sensitivity to small, dense
clumps.

The self-blocking transients seen in models B8 and B12 depend on the
resolution used; a resolution convergence study is needed to confirm
the effect. This effect is not seen in models A8 or A12. This could be
a result of the shallower density gradient or it could also be a
result of the resolution employed. We caution that a more complete
treatment of the radiative transfer could modify these results, hence
we do not make any quantitative claim about the frequency of
occurrence or lifetime of this morphology. We merely suggest that the
core-halo or extended emission morphology might arise from
self-blocking effects.  This is illustrated in figure 8, where we show
a contour plot of $n_e^2$ for model B12 together with a contour image
of G31.3$-$0.2, a UC~\hii region with extended emission.  The model does
not purport to explain the G31.3$-$0.2 region, which is about 30 times
larger, but it does suggest how its formation might occur.

\subsection{Summary}

To summarize, the models indicate that UC~\hii regions are long-lived
objects while they are inside their parental cloud cores. When stars
escape from these cores, transitions to larger structures, possibly
UC~\hii regions with extended emission, are expected to occur.
Conversely, stars can also fall back into the core owing to the
gravitational potential of the cloud. The more extended ionized gas
would recombine, leaving smaller ionized structures.

We thank Michael L. Norman and the Laboratory for Computational
Astrophysics for the use of ZEUS-3D. G.G.-S. thanks the 1998 OAN
summer students for the mosaic of M16 obtained with the 84 cm
telescope at San Pedro M\'artir.  The computations were performed on
the SGI Origin 2000 at IA-UNAM (Ensenada). This work has been
partially supported by grants from DGAPA-UNAM (IN130698, IN117799 \&
IN114199) and CONACyT (32214-E, 36568-E, \& E-43121).

\clearpage

\begin{figure}
\figurenum{1a}
\epsscale{0.8}\plotone{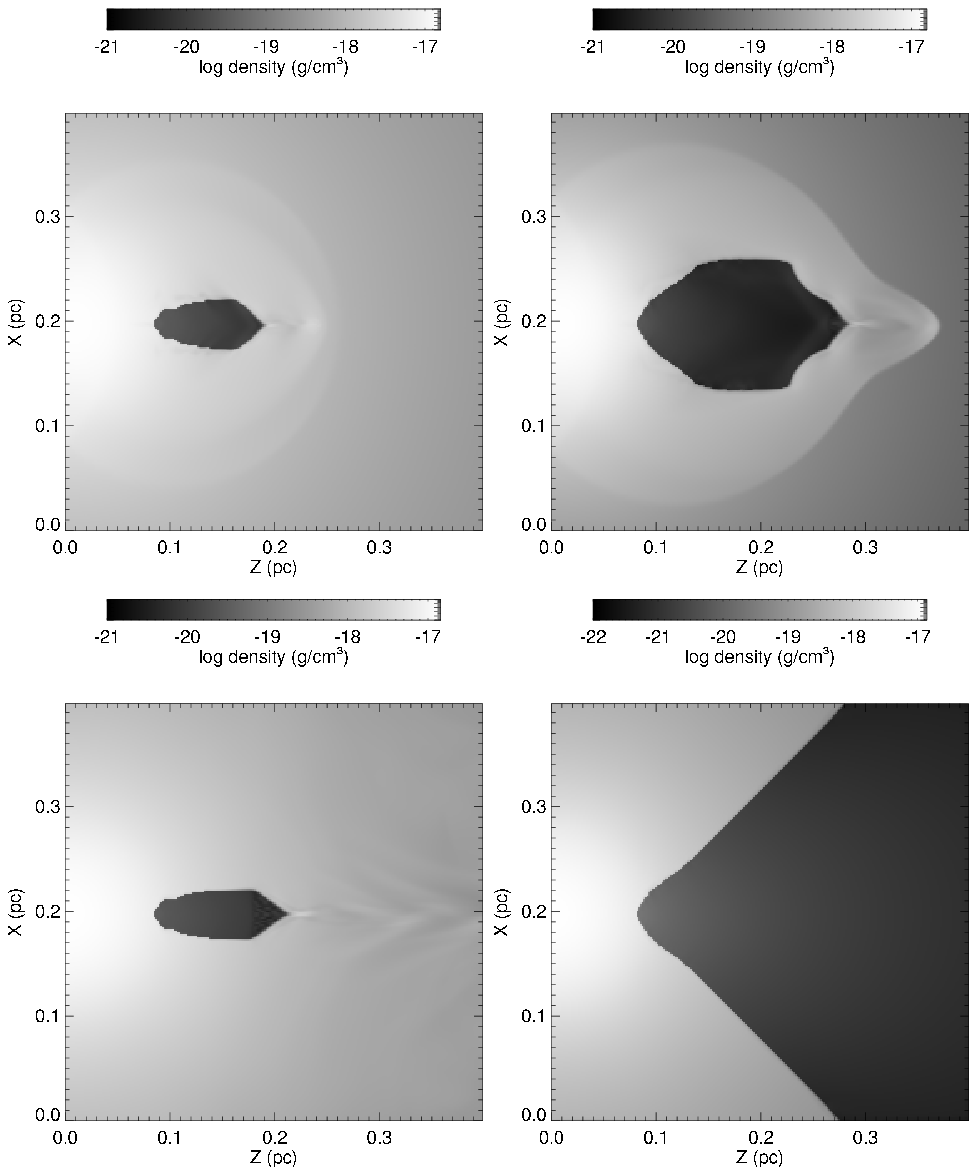}
\caption{Gas density snapshots of
   models A0 (left) and B0 (right). The star is located at the edge
   of the core, 0.1 pc from the core center: $[(x,z)_{\star} = (0.2,0.1)]$
   while  $[(x,z)_{core~center} = (0.2,0.0)]$.  
  Times in years are $t = 1 \times 10^5 $ (top), $t =
  6 \times 10^5 $ (bottom).  Model A0 reaches pressure equilibrium,
  while model B0 produces a blister.  For comparison, we note that a log
  density (g cm$^{-3}$) of $-19$ corresponds to number densities of 
  $3\times 10^4$~cm$^{-3}$ for molecular hydrogen
  and  $6 \times 10^4$~cm$^{-3}$ for ionized hydrogen.  
\label{fig1a}} 
\end{figure}
\clearpage

\begin{figure}
\figurenum{1b}
\epsscale{0.8}\plotone{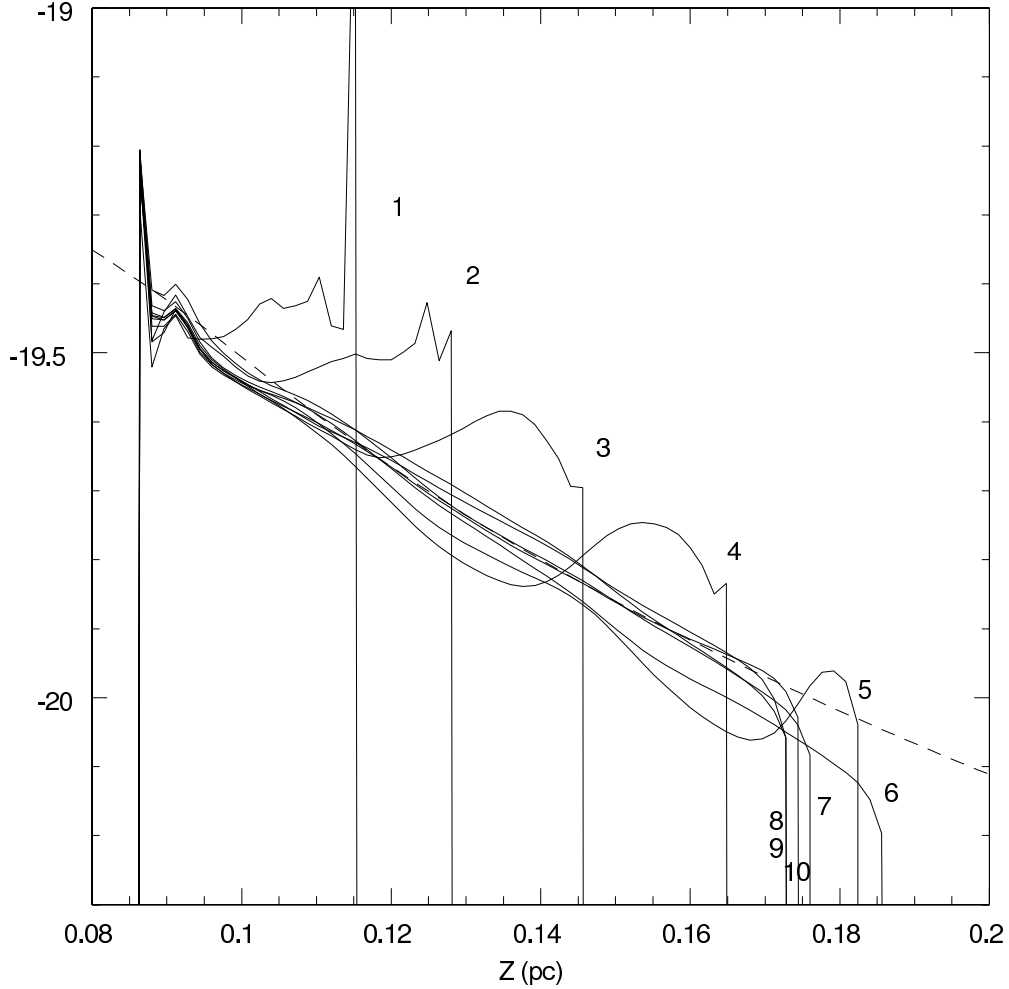}
\caption{Evolution of the photoionized gas density of model A0.
Curves are in labeled in units of 20,000 yr. That is, curves 1, 2, 3... 
correspond to times of 20, 40, 60... kilo years.
The star is located at Z = 0.1~pc.
The short dashed line represents the initial density distribution 
divided by 200, which is the expected equilibrium value 
($\rho \propto r^{-2}$ for $r \geq r_{\rm c}$).  The vertical axis shows
log mass density in units of g~cm$^{-3}$; values 
Log mass densities of $-19.5$ and $-20$~g~cm$^{-3}$ correspond to number 
densities of $1.9 \times 10^4$ and $6.0 \times 10^3$~cm$^{-3}$, respectively.
\label{fig1b}}
\end{figure}
\clearpage

\begin{figure}
\figurenum{1c}
\epsscale{0.8}\plotone{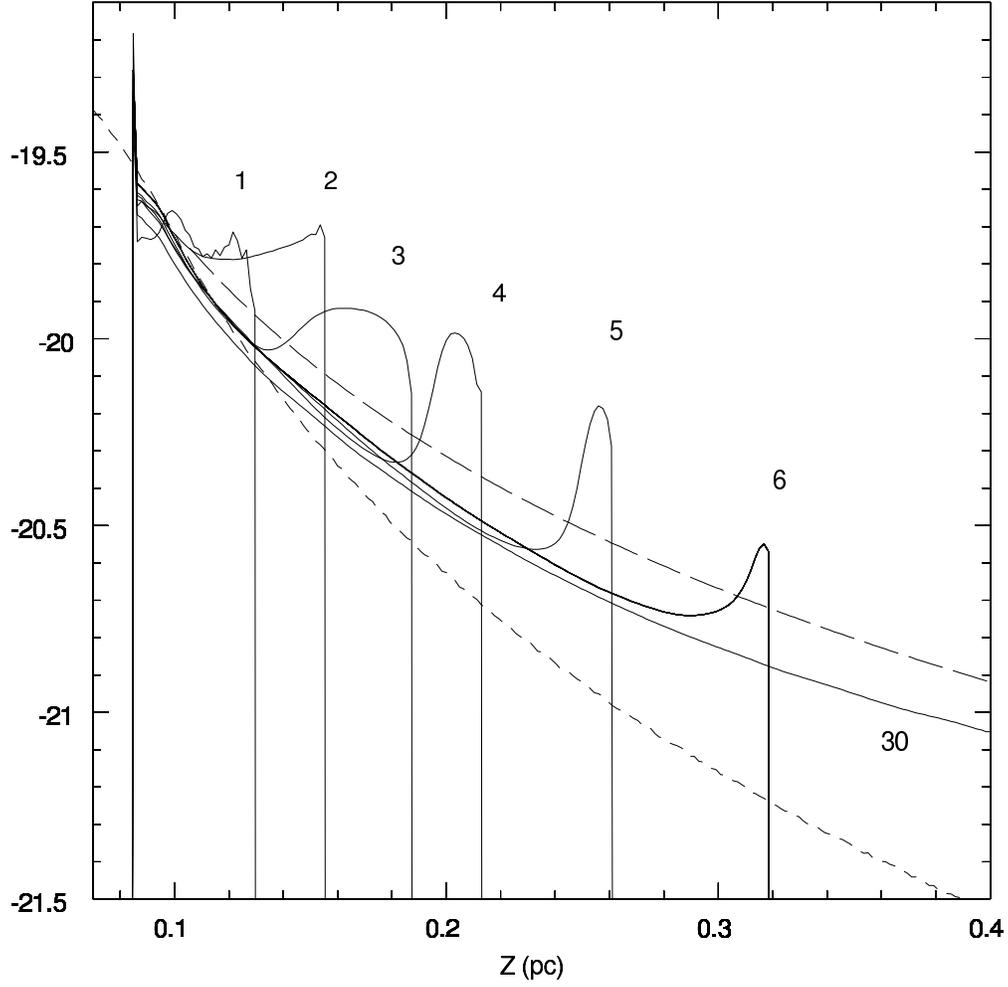}
\caption{Same as Figure 1b but for model B0.
The curves labeled 1, 2, 3... correspond to times in multiples of 20,000 yr.
The star is located at Z = 0.1~pc.
The short-dashed line represents the initial density structure divided by 
200, which is the expected equilibrium value ($\rho \propto r^{-3}$ for 
$r \geq r_{\rm c}$).
The long-dashed line is an $r^{-2}$ power law with the same central density.
Log mass densities of $-20$, $-20.5$ and $-21$~g~cm$^{-3}$ correspond to 
number densities of 6000, 1900, and 600~cm$^{-3}$, respectively.
\label{fig1c}}
\end{figure}
\clearpage

\begin{figure}
\figurenum{2} 
\epsscale{0.8}\plotone{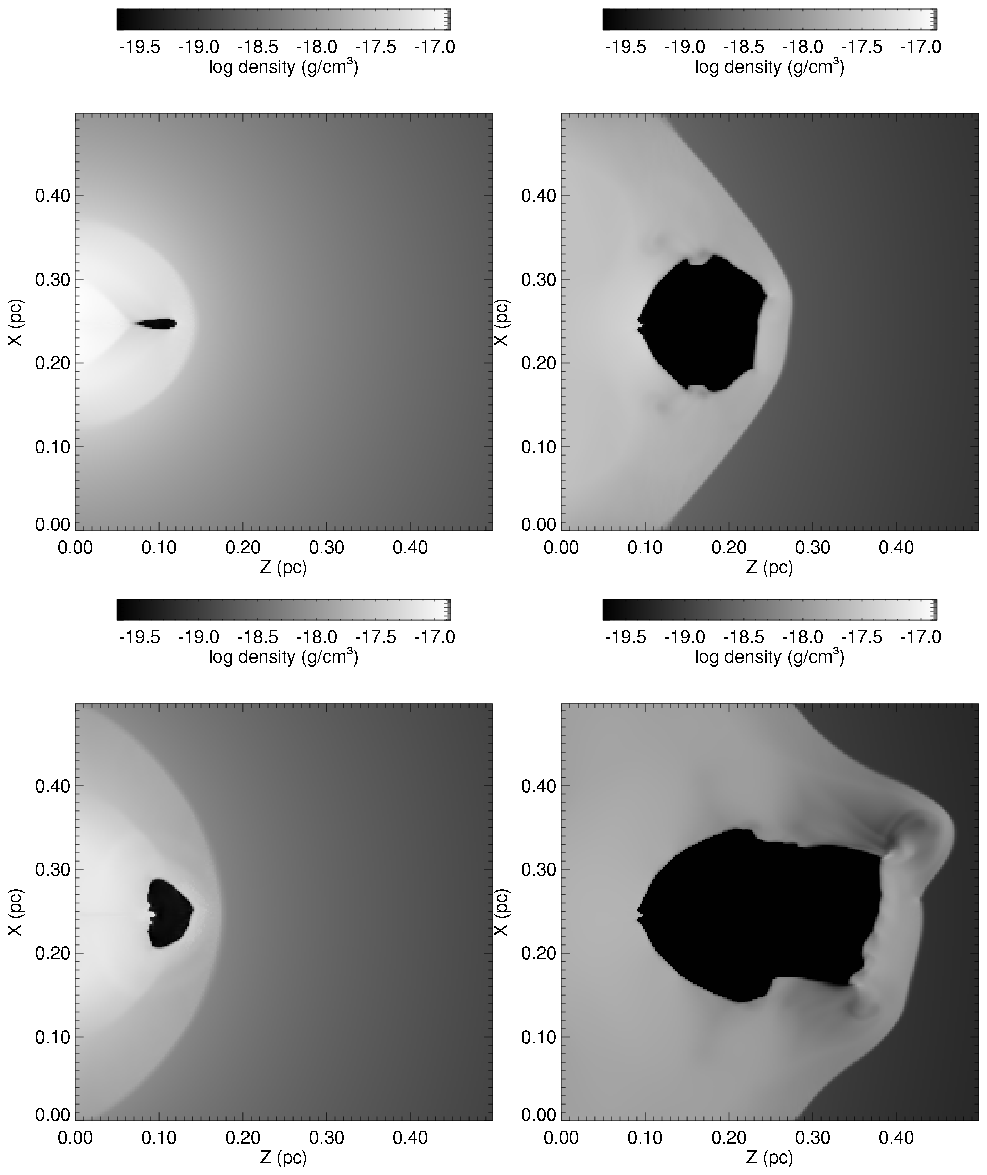}
\caption{Gas density snapshots of model A2.
Times in units of $10^5$ years are 0.8 (top left), 1.6 (bottom left), 
2.4 (top right), and 3.2 (bottom right).  The star departs from the core 
center at $t = 0$ with $v_{\star}=2 \kms$. 
X and Z coordinates are fixed to the star $ [(x,z)_{\star} = (0.25,0.1)]$,
but the final stellar position is 0.65 pc from the starting point.  This motion
is represented in the model by the molecular core moving leftward, out of the figure.
Log mass densities of $-17$, $-17.5$ and $-18$~g~cm$^{-3}$ correspond to 
number densities of $3 \times 10^6$, $9.5 \times 10^5$, and 
$3 \times 10^5$~cm$^{-3}$, respectively, for the molecular gas.
\label{fig2}}
\end{figure}
\clearpage

\begin{figure} 
\figurenum{3}
\epsscale{0.8}\plotone{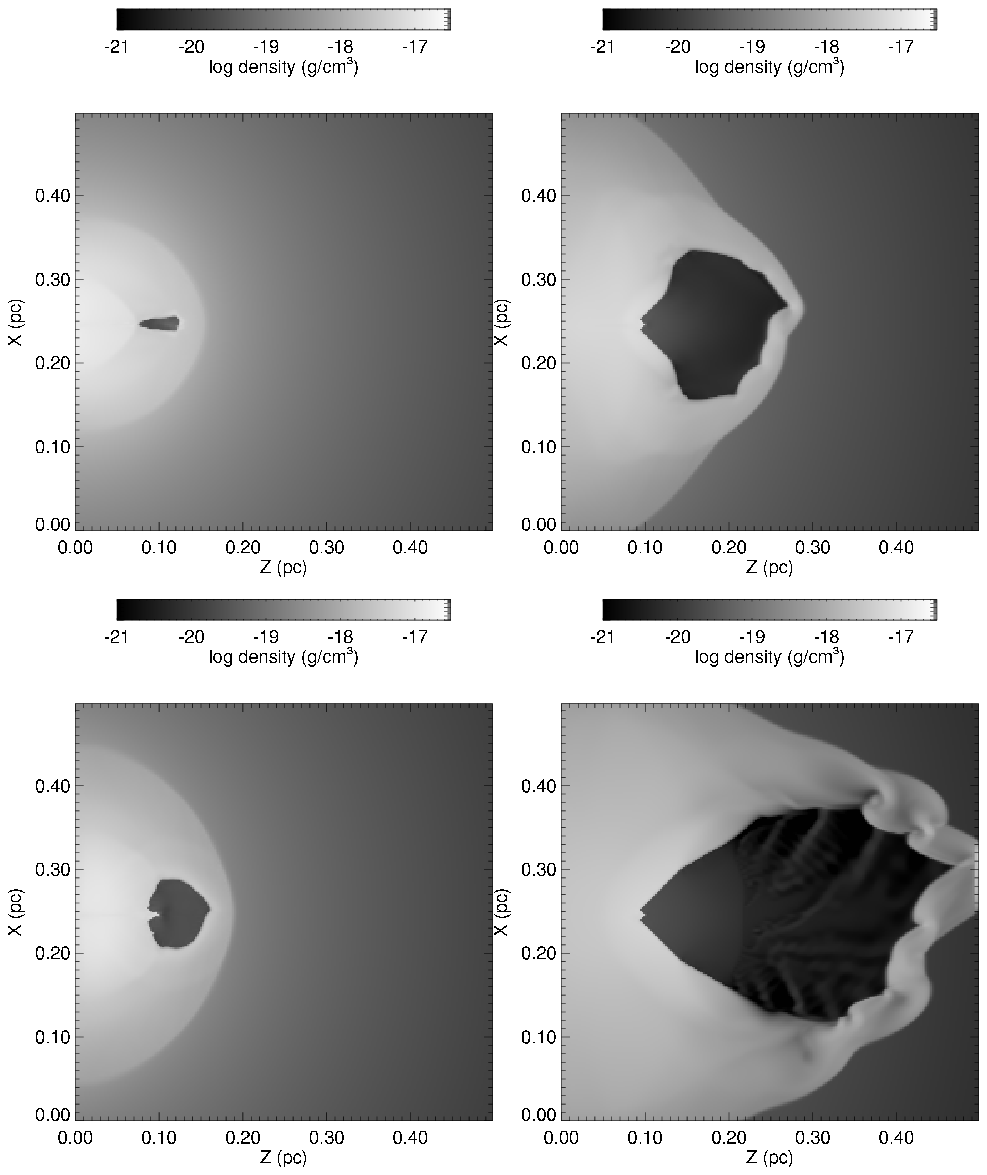}
\caption{Gas density snapshots of model B2.
Times in units of $10^5$ years are 0.8 (top left), 1.2 (bottom left),
1.66 (top right), and 2 (bottom right).  The star departs from the core
center at $t=0$ with $v_{\star}=2 \kms$.
X and Z coordinates are fixed to the star $[(x,z)_{\star} = (0.25,0.1)]$.
The final stellar position is 0.4 pc from the starting point.  Because of the
steeper density gradient, the expansion of the HII region in this model is more
rapid than that shown in Figure 2; the final frame of Figure 3 is intermediate
in time between the second and third frames of Figure 2.
\label{fig3}}
\end{figure}
\clearpage

\begin{figure}
\figurenum{4} 
\epsscale{0.8}\plotone{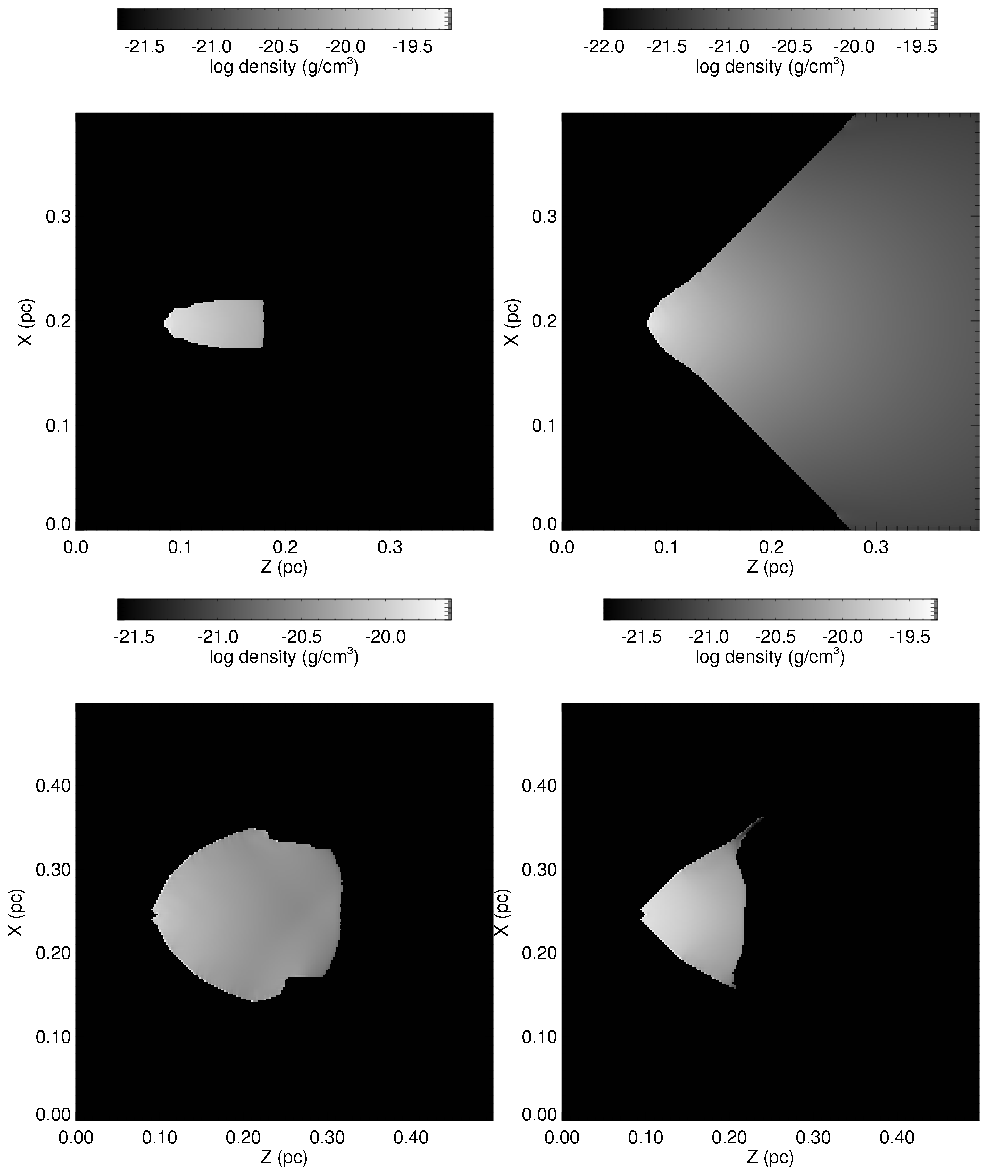}
\caption{Final photoionized gas densities for models
A0 (top left), B0 (top right), A2 (bottom left) and B2 (bottom right).
Times correspond to the end times in Figure 1a ($6 \times 10^5$~yr
for A0 and B0), Figure 2 ($3.2 \times 10^5$~yr for A2), and Figure 3 
($2.0 \times 10^5$~yr for B2).  A log mass density of $-20$ corresponds
to an electron density of 6,000~cm$^{-3}$ for the ionized gas.
Black areas show molecular gas and do not follow the log density gray-scale.
\label{fig4}}
\end{figure}
\clearpage

\begin{figure} 
\figurenum{5a}
\caption{Total gas density snapshots of models A8 (left panels) and A12 
(right panels).
Times in units of $10^5$ years are (top to bottom): 
0.12, 0.4, 0.64, 0.88, 1.12, 1.4, 1.64, 1.88  (for both models).
The cloud core center is at $[(x,z)_{core} = (0.1,0.1)]$ on the first frame,
and the simulation
begins with the star on the left edge of the core, at initial position
$[(x,z)_{\star} = (-0.1,0.1)]$.  As the simulation evolves, the star first
moves toward the core center with $v_{\star}$ = 8  and 12~\kms, respectively.
Eventually the star passes through the core and exits into the density 
ramp on the $+z$ side of the core.
The final $X-Z$ stellar position corresponds to 1.535 pc and 2.3 pc, 
respectively, from the cloud center $[(x,z)_{\star} = (0.1,0.1)]$.
\label{fig5a}}
\end{figure}
\clearpage
 
\begin{figure}
\figurenum{5b} 
\caption{Photoionized gas density counterpart of Figure 5a.
\label{fig5b}}
\end{figure}
\clearpage

\begin{figure}
\figurenum{6a} 
\caption{Total gas density snapshots of models B8 (left panels) and B12
(right panels).
Times in units of $10^5$ years are (top to bottom):
0.12, 0.28, 0.4, 0.52, 0.64, 0.76, 0.88, 1  (for both models).
The cloud core center is at $[(x,z)_{core} = (0.1,0.1)]$ on the first frame,
and the simulation
begins with the star on the left edge of the core, at initial position
$[(x,z)_{\star} = (-0.1,0.1)]$.  As the simulation evolves, the star first
moves toward the core center with $v_{\star}$ = 8  and 12~\kms, respectively.
Eventually the star passes through the core and exits into the density 
ramp on the $+z$ side of the core.
The final stellar position is 0.82  pc and 1.23 pc, respectively,
from the starting position.
\label{fig6a}}
\end{figure}
\clearpage

\begin{figure}
\figurenum{6b}  
\caption{Photoionized gas density counterpart of Figure 6a.
Frames 4 (bottom of first column) and 11 (third frame of third column)
illustrate the effect of the star overtaking a leading, piled-up clump
of gas.  This could result in either a core-halo morphology or an 
ultracompact HII region with extended emission.  See \S 5.4 and figure 8.
\label{fig6b}}
\end{figure}
\clearpage

\begin{figure}
\figurenum{7}   
\caption{Comparison of the famous central part of M16 (top) 
  in H$\alpha$ with model B8 at $ t = 8.8 \times 10^4 $ yr (bottom).
 The bottom left frame shows the ionized gas density while the bottom 
 right frame shows 
 the total gas density.  Although the model does not pretend to reproduce 
 M16, the similarity is striking.}
\label{fig7}
\end{figure}
\clearpage

\begin{figure}
\figurenum{8}   
\epsscale{1.0}\plotone{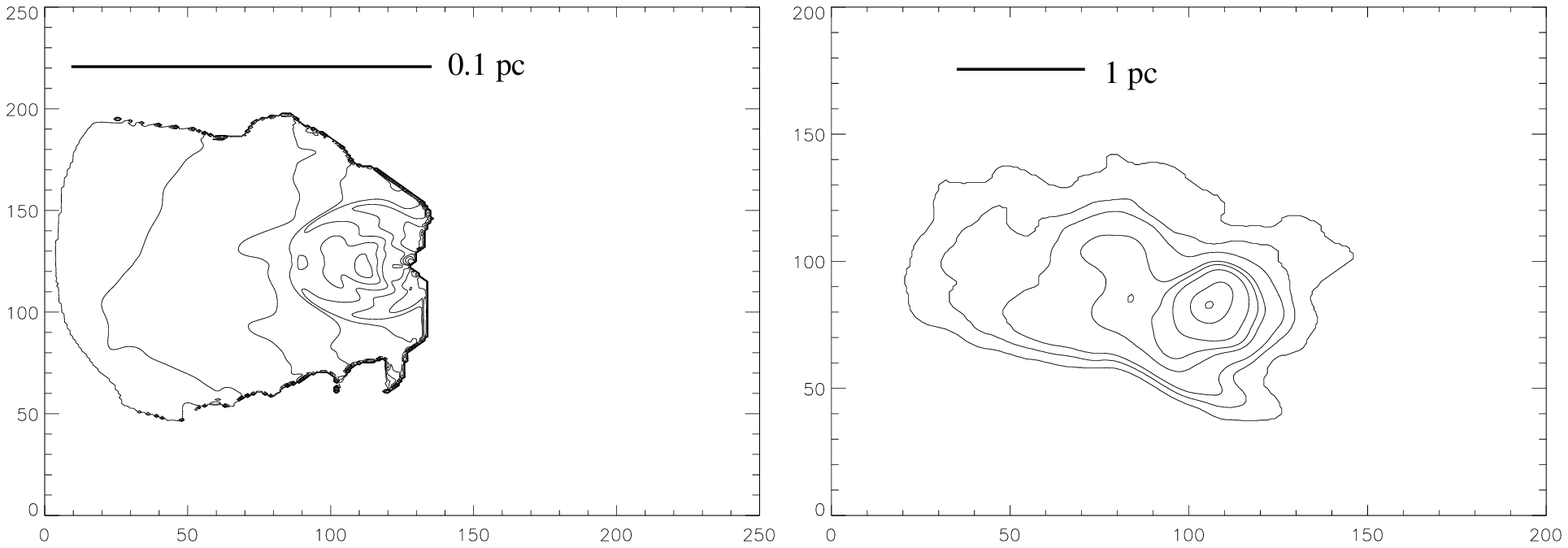}
\caption{Comparison of the ionized gas density (squared) of model B12 with an image
 of G31.3$-$0.2.  The contours of the left frame are proportional to $n_e^2$
 which in turn is proportional to the radio flux density in the case of optically
 thin emission.  The right frame (adapted from Kurtz et al. 1999) shows the
 radio emission of G31.3$-$0.2, which has extended, lower density gas 
 surrounding a compact, higher density component.  The 1 pc size scale assumes a 
 distance of 5.7~kpc.
 The model does not pretend to reproduce the detailed structure of G31.3$-$0.2, but rather
 is intended to show that the self-blocking effect can give rise, at least on small scales,
 to the morphology of a UC~\hii region with extended emission.}
\label{fig8}
\end{figure}
\clearpage

\begin{deluxetable}{lccccc}
\tablecaption{Model Attributes\label{tbl-1}}
\tablewidth{0pt}
\tablehead{
\colhead{Model} & \colhead{$V_\star$} & \colhead{Halo Density} & \colhead{Grid scale} &
\colhead{Lifetime as} & \colhead{Lifetime as Cometary}  \\
  &  \colhead {\kms} & \colhead{Gradient} & \colhead{pc} & \colhead{UC HII ($10^5$ yr)}
  & \colhead{UC HII ($10^4$ yr)} }
\startdata
 A0         & 0  & $r^{-2}$ & 0.4 & $\infty$ &  0   \\
 B0         & 0  & $r^{-3}$ & 0.4 & 0.8      &  0   \\
 A2         & 2  & $r^{-2}$ & 0.5 & 2.0      &  8   \\
 B2         & 2  & $r^{-3}$ & 0.5 & 1.4      &  8   \\
 A8         & 8  & $r^{-2}$ & 0.2 & 0.64     &  6.4 \\
 B8         & 8  & $r^{-3}$ & 0.2 & 0.4      &  4   \\
 A12        & 12 & $r^{-2}$ & 0.2 & 0.64     &  6.4 \\
 B12        & 12 & $r^{-3}$ & 0.2 & 0.4      &  4   \\
\enddata
\end{deluxetable}
\clearpage

\end{document}